\documentclass[aps,prl,twocolumn,floatfix,superscriptaddress]{revtex4-1}
\pdfoutput=1
\usepackage{graphicx}
\usepackage{amsmath,amsfonts,amssymb}
\usepackage[colorlinks=true,linkcolor=blue,citecolor=blue]{hyperref}
\usepackage{siunitx}
\usepackage{color,soul}

\begin{document}
\title{Mobility enhancement in graphene by in situ reduction of random strain fluctuations}

\author{Lujun Wang}
\email{lujun.wang@unibas.ch}
\affiliation{Department of Physics, University of Basel, Klingelbergstrasse 82, CH-4056 Basel, Switzerland}
\affiliation{Swiss Nanoscience Institute, University of Basel, Klingelbergstrasse 82, CH-4056 Basel, Switzerland}

\author{P\'eter Makk}
\email{peter.makk@unibas.ch}
\affiliation{Department of Physics, University of Basel, Klingelbergstrasse 82, CH-4056 Basel, Switzerland}
\affiliation{Department of Physics, Budapest University of Technology and Economics and Nanoelectronics Momentum Research Group of the Hungarian Academy of Sciences, Budafoki ut 8, 1111 Budapest, Hungary}

\author{Simon Zihlmann}
\affiliation{Department of Physics, University of Basel, Klingelbergstrasse 82, CH-4056 Basel, Switzerland}

\author{Andreas Baumgartner}
\affiliation{Department of Physics, University of Basel, Klingelbergstrasse 82, CH-4056 Basel, Switzerland}
\affiliation{Swiss Nanoscience Institute, University of Basel, Klingelbergstrasse 82, CH-4056 Basel, Switzerland}

\author{David I. Indolese}
\affiliation{Department of Physics, University of Basel, Klingelbergstrasse 82, CH-4056 Basel, Switzerland}

\author{Kenji Watanabe}
\affiliation{National Institute for Material Science, 1-1 Namiki, Tsukuba, 305-0044, Japan}

\author{Takashi Taniguchi}
\affiliation{National Institute for Material Science, 1-1 Namiki, Tsukuba, 305-0044, Japan}

\author{Christian Sch\"onenberger}
\affiliation{Department of Physics, University of Basel, Klingelbergstrasse 82, CH-4056 Basel, Switzerland}
\affiliation{Swiss Nanoscience Institute, University of Basel, Klingelbergstrasse 82, CH-4056 Basel, Switzerland}

\begin{abstract}
Microscopic corrugations are ubiquitous in graphene even when placed on atomically flat substrates. These result in random local strain fluctuations limiting the carrier mobility of high quality hBN-supported graphene devices. We present transport measurements in hBN-encapsulated devices where such strain fluctuations can be in situ reduced by increasing the average uniaxial strain. When $\sim0.2\%$ of uniaxial strain is applied to the graphene, an enhancement of the carrier mobility by $\sim35\%$ is observed while the residual doping reduces by $\sim39\%$. We demonstrate a strong correlation between the mobility and the residual doping, from which we conclude that random local strain fluctuations are the dominant source of disorder limiting the mobility in these devices. Our findings are also supported by Raman spectroscopy measurements.


\end{abstract}

\maketitle

In the first generation of graphene devices, where $\mathrm{SiO_2}$ was used as the substrate, it is commonly believed that random charged impurities at the substrate surface are the dominant source of disorder limiting the device quality~\cite{Ando2006,Nomura2007,Hwang2007,Adam2007,Chen2008,Chen2008a,Hong2009}. One way to improve the device quality is to suspend graphene to spatially separate it from the charge traps~\cite{Bolotin2008,Du2008,Bolotin2008a,Tombros2011,Maurand2014}. Nowadays, a more widely used technique is to place graphene on hexagonal boron nitride (hBN)~\cite{Dean2010,Zomer2011,Mayorov2011,Wang2013}, which is atomically flat and expected to be free of surface charge traps. A significant improvement in device quality has been achieved, exhibiting very high carrier mobilities, enabling the observation of a series of new physical phenomena, such as the fractional quantum Hall effect \cite{Bolotin2009,Du2009,Dean2011}, transverse magnetic focusing~\cite{Taychatanapat2013,Lee2016,Chen2016} and various moir\'e superlattice effects~\cite{Ponomarenko2013,Dean2013,Hunt2013,Wang2019}. Although the mobility of hBN-supported graphene devices is generally higher than that of the $\mathrm{SiO_2}$-supported, the reported mobility values vary over a large range, suggesting another mechanism that limits the mobility. It has been pointed out that random strain fluctuations (RSFs) in graphene could be a dominant source of disorder leading to electron scattering~\cite{Katsnelson2008}. In a recent statistical study of many devices on hBN substrates, a clear correlation between the carrier mobility $\mu$ and the residual doping $n_0$ was found, pointing to RSFs as the dominant microscopic source of scattering~\cite{Couto2014}. The residual doping caused by charge fluctuations manifests in a broadening of the resistance peak around the charge neutrality point (CNP). Similar results have been found as well in bilayer graphene~\cite{Engels2014}. 


Ripples and pronounced corrugations can form naturally in graphene due to its two-dimensional nature, as, for example, demonstrated by transmission electron microscopy in suspended graphene membranes~\cite{Meyer2007}. In stacked layers, microscopic corrugations can spontaneously form during exfoliation due to thermal fluctuations at room temperature~\cite{Abedpour2007,Fasolino2007,Katsnelson2008}. These corrugations might persist through the fabrication processes and give rise to RSFs in the final device. In $\mathrm{SiO_2}$-supported devices, nanometer-scale ripples have been observed in scanning probe microscopy studies~\cite{Ishigami2007,Stolyarova2007,Geringer2009,Cullen2010} and their effects on electron transport have been reported in weak localization studies~\cite{Morozov2006,Tikhonenko2008,Lundeberg2010}. Although the hBN surface is typically much flatter, height fluctuations are still present in hBN-supported graphene devices~\cite{Dean2010}, which can result in RSFs. These RSFs have been confirmed in Raman spectroscopy measurements~\cite{Neumann2015,Banszerus2017}.



\begin{figure}[htb]
	\centering
	\includegraphics[]{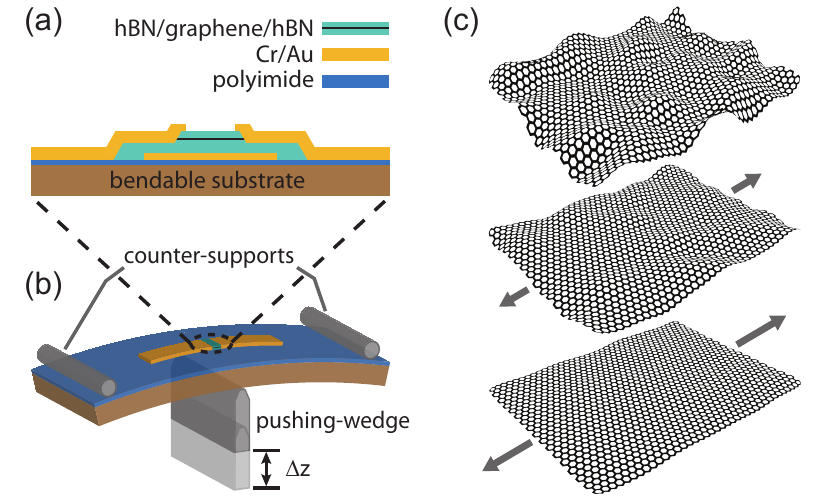}
	\caption{Schematics of \textbf{(a)} the device cross section and \textbf{(b)} the three-point bending setup. The bending of the substrate is determined by the displacement of the pushing-wedge, $\mathrm{\Delta}z$. \textbf{(c)} Illustration of the effects of reducing the strain fluctuations. The arrows indicate the direction and the strength of the externally induced strain by substrate bending mediated by contacts.}
	\label{fig:fig1}
\end{figure}

Here we demonstrate in a direct experiment that RSFs can be the mechanism limiting the mobility of encapsulated devices. We compare the transport characteristics of individual devices before and after increasing the average uniaxial strain, which directly reduces the strain fluctuations in the same device. In Fig.~\ref{fig:fig1}(c) the RSFs in graphene lattice are illustrated, which we believe can be reduced gradually by increasing the average strain, as indicated by the arrows. The reduction of the RSFs due to increasing average strain is further confirmed by directly probing the RSFs using Raman spectroscopy~\cite{Neumann2015}. This not only allows us to determine the dominant microscopic mechanism, but also to actually increase the mobility of the device. 

\begin{figure}[htb]
	\centering
	\includegraphics[]{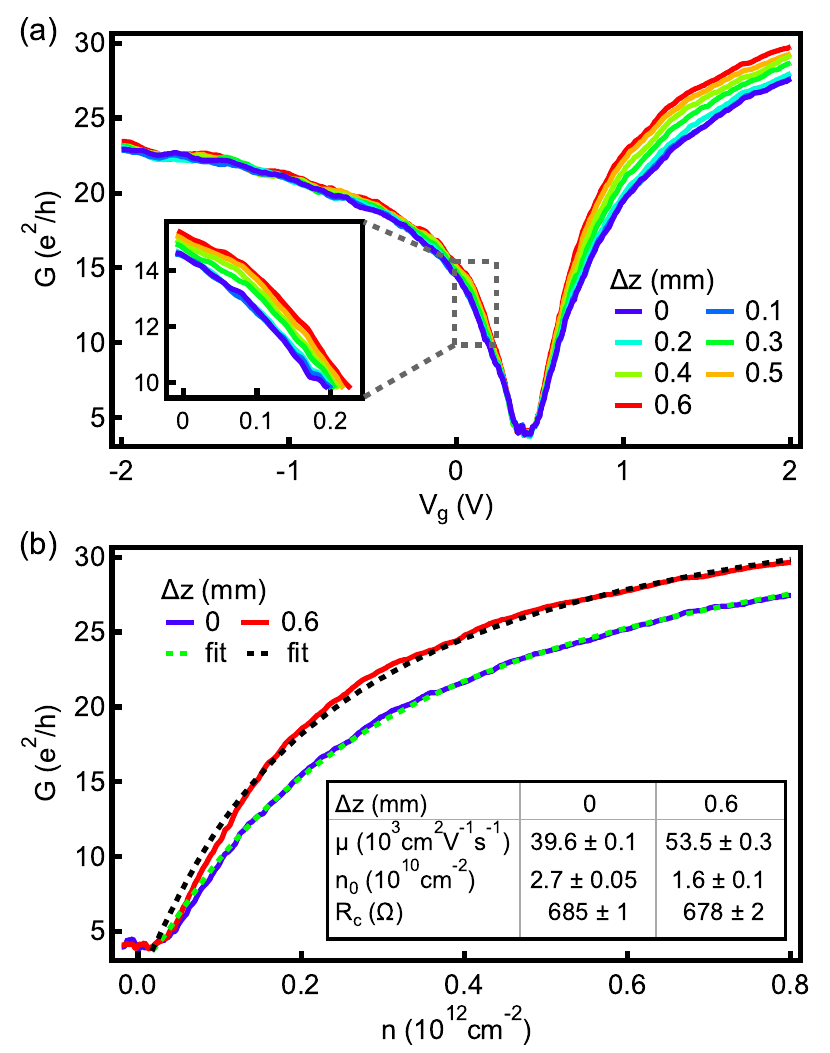}
	\caption{\textbf{(a)} Two-terminal differential conductance $G$ plotted as a function of gate voltage $V_\mathrm{g}$ for different $\mathrm{\Delta}z$ values. The slope of the curves becomes steeper for larger $\mathrm{\Delta}z$, for both the electron and hole side. The inset shows a zoom-in to the hole side. \textbf{(b)} $G$ versus $n$ for two different $\mathrm{\Delta}z$ on the electron side. The fits according to Eq.~\ref{eq:drude} are shown as dashed lines for $\mathrm{\Delta}z = 0$ and \SI{0.6}{\milli\meter}, respectively, with the fitting parameters given in the table.}
	\label{fig:fig2}
\end{figure}


The setup of the experiment is shown schematically in Fig.~\ref{fig:fig1}(a,b). It allows us to tune the average uniaxial strain in hBN-encapsulated graphene devices by bending a flexible substrate~\cite{Wang2019a}. The displacement $\mathrm{\Delta}z$ of the pushing-wedge relative to the mounting position determines the deformation of the substrate and is used to tune the average strain in the graphene. The devices are fabricated using a dry-transfer method~\cite{Wang2013}, where we pick up a \SI{\sim20}{\nano\meter} thick hBN as the top layer, then an exfoliated monolayer graphene flake from natural graphite and a \SI{\sim30}{\nano\meter} thick hBN as the bottom layer. The assembled stack is then deposited onto a metallic gate structure prefabricated on a polyimide-coated phosphor bronze plate. Edge contacts~\cite{Wang2013} (Cr/Au, \SI{5}{\nano\meter}/\SI{110}{\nano\meter}) are made with a controlled etching recipe, which stops in the middle of the bottom hBN, with the remaining hBN acting as the insulating layer between the contacts and the bottom gate~\cite{Wang2019a}, see Fig.~\ref{fig:fig1}(a). 


To investigate the effects of average strain on the transport characteristics of graphene, we measure the two-terminal differential conductance $G$ as a function of the gate voltage $V_\mathrm{g}$ for different $\mathrm{\Delta}z$ values, as plotted in Fig.~\ref{fig:fig2}(a). The measurements were performed at low temperature ($T = \SI{4.2}{\kelvin}$) using standard low-frequency lock-in techniques. The CNP is at $V_\mathrm{g}=\SI{0.4}{\volt}$, indicating an offset p-doping in our device. The conductance of the graphene increases faster when gated away from the CNP for larger $\mathrm{\Delta}z$, suggesting an increase in field effect mobility with increasing $\mathrm{\Delta}z$. This effect is reversible when $\mathrm{\Delta}z$ is decreased (see Supplementary Material). A displacement of $\mathrm{\Delta}z = \SI{0.6}{\milli\meter}$ corresponds to $\sim 0.2\%$ of average strain, which is determined from Raman measurements shown later~\cite{Wang2019a}. The conductance starts to saturate at higher gate voltages because of the contact resistance. On the hole side (p-doping), a p-n junction forms near each contact due to the n-doping from the contact, resulting in a sightly larger contact resistance and a lower saturation conductance, which renders the mobility-change less visible. The zoomed-in data in the inset of Fig.~\ref{fig:fig2}(a) shows qualitatively the same effect as for the electron side.


To quantitatively evaluate the effects of strain tuning on the electrical properties of graphene, we fit each curve on the electron side (n-doping) with the following formula based on the Drude model~\cite{Hong2009,Dean2010}: 
\begin{equation}
	G = \frac{1}{\frac{\alpha }{e\mu \sqrt{n^{2} + n_{0}^{2}}}+R_\mathrm{c}},
	\label{eq:drude}
\end{equation}
where $e$ is the elementary charge and $\alpha$ is the geometry factor describing the aspect ratio, which is 1.28 in this case (see Supplementary Material). The fitting parameters are the charge-carrier density independent mobility $\mu$, the residual doping $n_{0}$ around the CNP and the contact resistance $R_\mathrm{c}$. The charge-carrier density $n$ is calculated from the applied gate voltage $V_\mathrm{g}$ with a lever arm of \SI{5.13e11}{\per\square\centi\meter\per\volt} using a parallel plate capacitor model. The thickness of the bottom hBN, which is the gate dielectric, is determined by atomic force microscopy. Two examples of the fitting are shown as dashed lines for $\mathrm{\Delta}z$ = 0 and \SI{0.6}{\milli\meter} in Fig.~\ref{fig:fig2}(b) with the corresponding parameters given in the inset.

\begin{figure}[htb]
	\centering
	\includegraphics[]{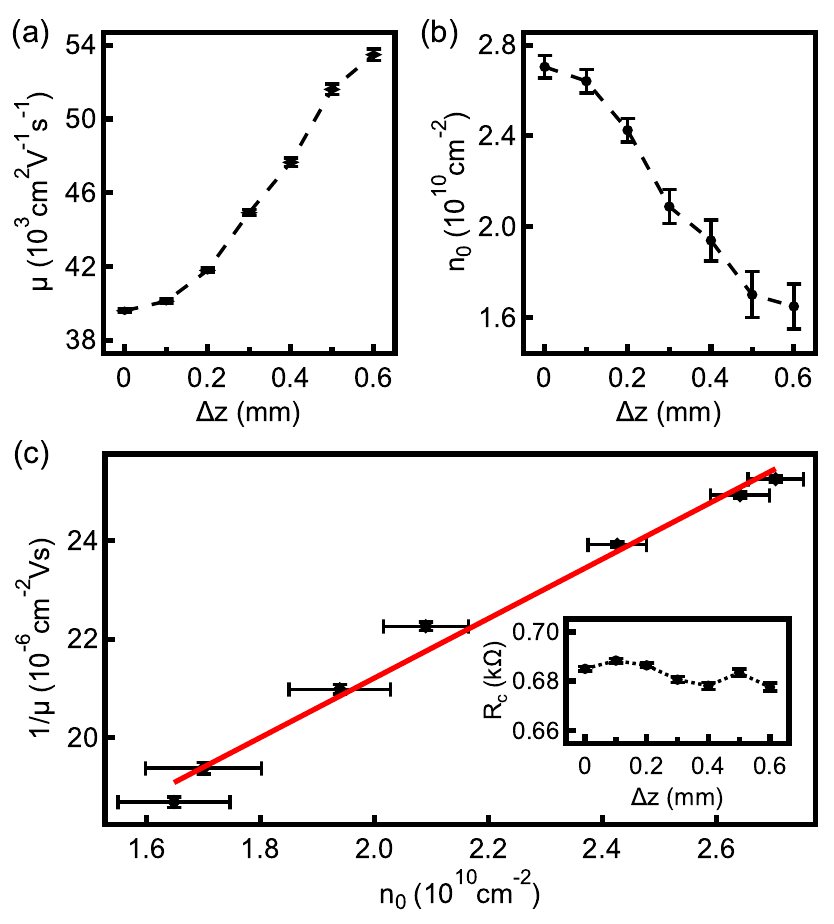}
	\caption{\textbf{(a)} Extracted field effect mobility $\mu$ and \textbf{(b)} residual doping $n_{0}$ values from fitting plotted as a function of $\mathrm{\Delta}z$ on the electron side. The error bars are the standard errors from fits. The mobility $\mu$ shows an increase with increasing $\mathrm{\Delta}z$ while $n_{0}$ shows a decrease. \textbf{(c)} Data of \textbf{(a)} and \textbf{(b)} plotted as $1/\mu$ versus $n_{0}$, showing a clear linear relation. The red line is a linear fit to the data with $1/\mu = (0.146\pm0.007) \times (h/e)n_{0}+1/\mu_{0}$ and $\mu_{0} \approx \SI{110000}{\square\centi\meter\per\volt\per\second}$. The inset shows the extracted contact resistance $R_\mathrm{c}$ (including \SI{\sim350}{\ohm} line resistance) for different $\mathrm{\Delta}z$.}
	\label{fig:fig3}
\end{figure}

The fitting results for $\mu$ and $n_{0}$ are plotted as a function of $\mathrm{\Delta}z$ in Fig.~\ref{fig:fig3}(a,b), respectively. The mobility $\mu$ shows a clear increase with increasing $\mathrm{\Delta}z$, while $n_{0}$ decreases significantly. The change is slower in the beginning, which might be attributed to a small mechanical hysteresis of the bending setup. The extracted contact resistance $R_\mathrm{c}$ (including \SI{\sim350}{\ohm} line resistance) is shown in the inset of Fig.~\ref{fig:fig3}(c) and is essentially unaffected by the bending, demonstrating the mechanical robustness of the device for these levels of applied average strain~\cite{Wang2019a}. The mobility increases from \SI{\sim40000}{\square\centi\meter\per\volt\per\second} to \SI{\sim54000}{\square\centi\meter\per\volt\per\second} when $\mathrm{\Delta}z$ is increased from 0 to \SI{0.6}{\milli\meter}. At the same time the residual doping drops gradually from \SI{\sim2.7e10}{\per\square\centi\meter} ($\mathrm{\Delta}z = 0$) to \SI{\sim1.6e10}{\per\square\centi\meter\square} ($\mathrm{\Delta}z = \SI{0.6}{\milli\meter}$). The \mbox{($\mu$, $n_0$)} pairs are plotted as $1/\mu$ versus $n_{0}$ in Fig.~\ref{fig:fig3}(c), clearly demonstrating the proportionality between $1/\mu$ and $n_0$. The same analysis is performed for the hole side and similar results are obtained with a larger contact resistance (see Supplementary Material), which is consistent with the interpretation that the p-n junction makes the effect less pronounced on the hole side.

Since the graphene is encapsulated with hBN, it is very unlikely that the small applied average strain changes the charged impurities at the graphene-hBN interfaces, ruling them out as dominant mechanism for the observed mobility increase. An artificial effect due to the change of the gate capacitance with strain is also ruled out~\cite{Wang2019a}, because the CNP appears at the same gate voltage for all $\mathrm{\Delta}z$ values.

RSFs have been identified theoretically as a possible source of disorder limiting charge carrier mobility~\cite{Katsnelson2008}. Strong evidence of this mechanism has been found in a statistical study involving many devices, where a clear linear relation between $1/\mu$ and $n_0$ was observed, with $1/\mu \approx 0.118 \times (h/e)n_{0}$~\cite{Couto2014}. Moreover, a detailed microscopic mechanism was proposed in which the variation of $n_{0}$ was attributed to RSFs-induced scalar potentials, while the limitation in $\mu$ was attributed to randomly varying pseudomagnetic fields~\cite{Couto2014}. Fitting our data linearly in Fig.~\ref{fig:fig3}(c) yields $1/\mu = (0.146\pm0.007) \times (h/e)n_{0}+1/\mu_{0}$ and $\mu_{0} \approx \SI{110000}{\square\centi\meter\per\volt\per\second}$. It shows a similar slope ($\sim0.146 \times (h/e)$), allowing us to draw two conclusions. First, the charge carrier mobility is limited by RSFs and second, the control of the average strain allows us to control the RSFs and hence the mobility. The offset $1/\mu_{0}$ might imply another mobility limiting mechanism when RSFs are not dominating anymore. The value $\mu_{0} \approx \SI{110000}{\square\centi\meter\per\volt\per\second}$ nearly coincides with the mobility of the devices, in which no mobility enhancement due to increasing average strain is observed (discussed later).

Theoretically both, in-plane and out-of-plane, strain fluctuations can contribute to this effect~\cite{Couto2014}. In a previous study of weak localization on $\mathrm{SiO_2}$-supported graphene devices~\cite{Lundeberg2010}, a reduction of the phase coherence time $\tau_{\phi}$ was found for an increasing in-plane magnetic field. It has been attributed to an enhanced dephasing rate due to a random vector potential generated by the in-plane magnetic field penetrating out-of-plane corrugations in the graphene layer. Similar effects have been observed in encapsulated devices~\cite{Zihlmann2018,Zihlmann2019}, strongly suggesting that out-of-plane corrugations are also present in encapsulated graphene. We therefore attribute the mobility increase in our experiment to the reducing of out-of-plane strain fluctuations, as illustrated in Fig.~\ref{fig:fig1}(c).





\begin{figure}[htb]
	\centering
	\includegraphics[]{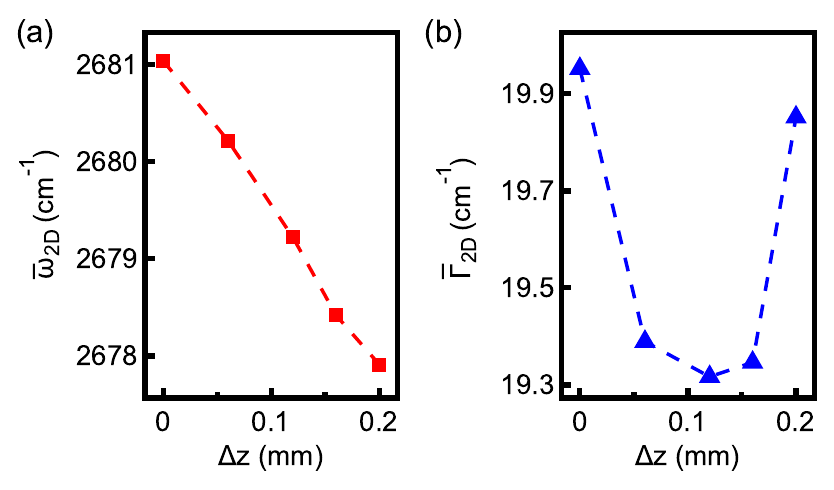}
	\caption{\textbf{(a)} Spatially averaged center frequency $\bar{\omega}_\mathrm{2D}$ of the Raman 2D peak for different $\mathrm{\Delta}z$, showing a linear decrease with increasing $\mathrm{\Delta}z$, which suggests an increasing average strain. \textbf{(b)} Spatially averaged linewidth $\bar{\Gamma}_\mathrm{2D}$ of the Raman 2D peak as a function of $\mathrm{\Delta}z$, exhibiting a nonmonotonic characteristics with a minimum of \SI{\sim19.3}{\per\centi\meter}.}
	\label{fig:fig4}
\end{figure}

To further substantiate our findings, we use spatially resolved Raman spectroscopy to directly probe the RSFs at room temperature. For small uniaxial strain, which is the case in our experiment, the graphene Raman 2D peak can be fitted by a single Lorentzian~\cite{Mohiuddin2009}, with a center frequency $\omega_\mathrm{2D}$ and linewidth $\Gamma_\mathrm{2D}$. The center frequency $\omega_\mathrm{2D}$ redshifts with increasing strain, while the linewidth $\Gamma_\mathrm{2D}$ broadens due to the splitting of the 2D peak~\cite{Mohr2009,Huang2010}. It has been shown that nanometer-scale strain inhomogeneities within the laser spot (\SI{\sim500}{\nano\meter}) also broadens the 2D peak~\cite{Neumann2015}, originating from averaging over regions with different local strain and hence different $\omega_\mathrm{2D}$. Therefore, $\Gamma_\mathrm{2D}$ can be used to probe the RSFs. We perform spatially resolved Raman spectroscopy and extract maps of $\omega_\mathrm{2D}$ and $\Gamma_\mathrm{2D}$ for different $\mathrm{\Delta}z$. The mean value of the center frequency $\bar{\omega}_\mathrm{2D}$ averaged over the whole device is plotted as a function of $\mathrm{\Delta}z$ in Fig.~\ref{fig:fig4}(a). It shifts linearly to lower values with increasing $\mathrm{\Delta}z$, indicating an increasing average strain in the graphene sheet~\cite{Mohiuddin2009}. The \SI{\sim3}{\per\centi\meter} shift at $\mathrm{\Delta}z = \SI{0.2}{\milli\meter}$ corresponds to an externally induced average strain of $\sim0.06\%$~\cite{Wang2019a}. In Fig.~\ref{fig:fig4}(b) the averaged value of the 2D peak linewidth $\bar{\Gamma}_\mathrm{2D}$ is plotted as a function of $\mathrm{\Delta}z$, showing nonmonotonic characteristics with a minimum of \SI{\sim19.3}{\per\centi\meter} at $\mathrm{\Delta}z = \SI{0.12}{\milli\meter}$. It first decreases with increasing $\mathrm{\Delta}z$ before increasing again, which can be explained by the competition between the two broadening mechanisms. The initial value of $\bar{\Gamma}_\mathrm{2D}$ (\SI{\sim20}{\per\centi\meter}) is larger than the intrinsic linewidth (\SI{\sim17}{\per\centi\meter}) of the 2D peak~\cite{Neumann2015}, indicating that RSFs are present in our graphene. We attribute the decrease of $\bar{\Gamma}_\mathrm{2D}$ to a reduction of the RSFs due to the externally applied strain, as illustrated in Fig.~\ref{fig:fig1}(c). When the broadening of the 2D peak induced by the increasing average strain dominates, $\bar{\Gamma}_\mathrm{2D}$ increases again with increasing $\mathrm{\Delta}z$. 




Our interpretation is also consistent with weak localization measurements we performed to extract characteristic scattering times (see Supplementary Material). We find that the intervalley scattering time $\tau_\mathrm{iv}$ is much longer than the elastic scattering time $\tau$ (determined from the mobility), implying that the mobility is not limited by intervalley scattering processes (scattering on short-range potentials, e.g. defects, edges). In contrast, the intravalley scattering time $\tau_\mathrm{*}$ (the time needed to break the effective single-valley time-reversal symmetry) is nearly identical to $\tau$, pointing to RSFs-induced random pseudomagnetic fields as the main factors limiting the mobility~\cite{Couto2014}. For charged impurities, it has been argued that $\tau_\mathrm{*}\gg\tau$~\cite{Couto2014}, which is not the case here.

We have observed a clear increase in the mobility with increasing average strain in more than 5 devices with their mobility values varying from \SI{\sim30000}{\square\centi\meter\per\volt\per\second} to \SI{\sim80000}{\square\centi\meter\per\volt\per\second}. In Fig.~\ref{fig:fig3}(a), there is also an indication that the mobility starts to saturate when it approaches higher values. For the devices with a mobility larger than \SI{\sim80000}{\square\centi\meter\per\volt\per\second}, the mobility-increase effect is absent. (Examples are presented in the Supplementary Material). These observations suggest that either the RSFs cannot be fully reduced by increasing the average strain or another mechanism is at play for ultra high mobility devices.

In conclusion, we have demonstrated an in situ reduction of the random RSFs in individual encapsulated graphene devices by increasing the average strain. In low-temperature transport measurements, an enhancement of the carrier mobility by $\sim35\%$ is observed while the residual doping reduces by $\sim39\%$ when $\sim0.2\%$ of average strain is applied to the graphene. The linear correlation between $1/\mu$ and $n_{0}$ reveals that random RSFs are the dominant scattering mechanism. These findings are further substantiated by Raman spectroscopy, in which the 2D peak linewidth $\Gamma_\mathrm{2D}$, first decreases with increasing average strain before the average strain induced broadening dominates. The in situ straining allows us to directly compare results on individual devices and to avoid statistics over different devices. Using this technique we have directly confirmed that random RSFs are the dominant scattering mechanism limiting the mobility in most hBN-supported graphene devices. For devices with even higher mobilities, either the reduction of RSFs is not possible, or another scattering mechanism becomes dominant.  

\subsection{Author contributions}
L.W. fabricated the devices, performed the measurements and did the data analysis. P.M. helped to understand the data. P.M., S.Z. and A.B. helped with the data analysis. P.M., S.Z. and D.I. supported the device fabrication. K.W. and T.T. provided high-quality hBN. C.S. initiated and supervised the project. L.W. wrote the paper and all authors discussed the results and worked on the manuscript. All data in this publication are available in numerical form at: \href{https://doi.org/10.5281/zenodo.3464530}{https://doi.org/10.5281/zenodo.3464530}.

\section*{Acknowledgments}
This work has received funding from the Swiss Nanoscience Institute (SNI), the ERC project TopSupra (787414), the European Union Horizon 2020 research and innovation programme under grant agreement No. 785219 (Graphene Flagship), the Swiss National Science Foundation, the Swiss NCCR QSIT, Topograph, ISpinText FlagERA network and from the OTKA FK-123894 grants. P.M. acknowledges support from the Bolyai Fellowship, the Marie Curie grant and the National Research, Development and Innovation Fund of Hungary within the Quantum Technology National Excellence Program (Project No. 2017-1.2.1-NKP-2017-00001). K.W. and T.T. acknowledge support from the Elemental Strategy Initiative conducted by the MEXT, Japan and the CREST (JPMJCR15F3), JST. The authors thank Francisco Guinea and Peter Rickhaus for fruitful discussions, and Sascha Martin and his team for their technical support.

\bibliography{MobilityChange}

%
%
%

\end{document}


\beginsupplement

\title{Supplementary Material for \\ Mobility enhancement in graphene by in situ reduction of random strain fluctuations}


\author{Lujun Wang}
\email{lujun.wang@unibas.ch}
\affiliation{Department of Physics, University of Basel, Klingelbergstrasse 82, CH-4056 Basel, Switzerland}
\affiliation{Swiss Nanoscience Institute, University of Basel, Klingelbergstrasse 82, CH-4056 Basel, Switzerland}

\author{P\'eter Makk}
\email{peter.makk@unibas.ch}
\affiliation{Department of Physics, University of Basel, Klingelbergstrasse 82, CH-4056 Basel, Switzerland}
\affiliation{Department of Physics, Budapest University of Technology and Economics and Nanoelectronics Momentum Research Group of the Hungarian Academy of Sciences, Budafoki ut 8, 1111 Budapest, Hungary}

\author{Simon Zihlmann}
\affiliation{Department of Physics, University of Basel, Klingelbergstrasse 82, CH-4056 Basel, Switzerland}

\author{Andreas Baumgartner}
\affiliation{Department of Physics, University of Basel, Klingelbergstrasse 82, CH-4056 Basel, Switzerland}
\affiliation{Swiss Nanoscience Institute, University of Basel, Klingelbergstrasse 82, CH-4056 Basel, Switzerland}

\author{David I. Indolese}
\affiliation{Department of Physics, University of Basel, Klingelbergstrasse 82, CH-4056 Basel, Switzerland}

\author{Kenji Watanabe}
\affiliation{National Institute for Material Science, 1-1 Namiki, Tsukuba, 305-0044, Japan}

\author{Takashi Taniguchi}
\affiliation{National Institute for Material Science, 1-1 Namiki, Tsukuba, 305-0044, Japan}

\author{Christian Sch\"onenberger}
\affiliation{Department of Physics, University of Basel, Klingelbergstrasse 82, CH-4056 Basel, Switzerland}
\affiliation{Swiss Nanoscience Institute, University of Basel, Klingelbergstrasse 82, CH-4056 Basel, Switzerland}

\maketitle


\section{Geometry factor and reversibility of the strain induced mobility increase}\label{sec:reversible}
\begin{figure}[H]
    \centering
      \includegraphics[width=\columnwidth]{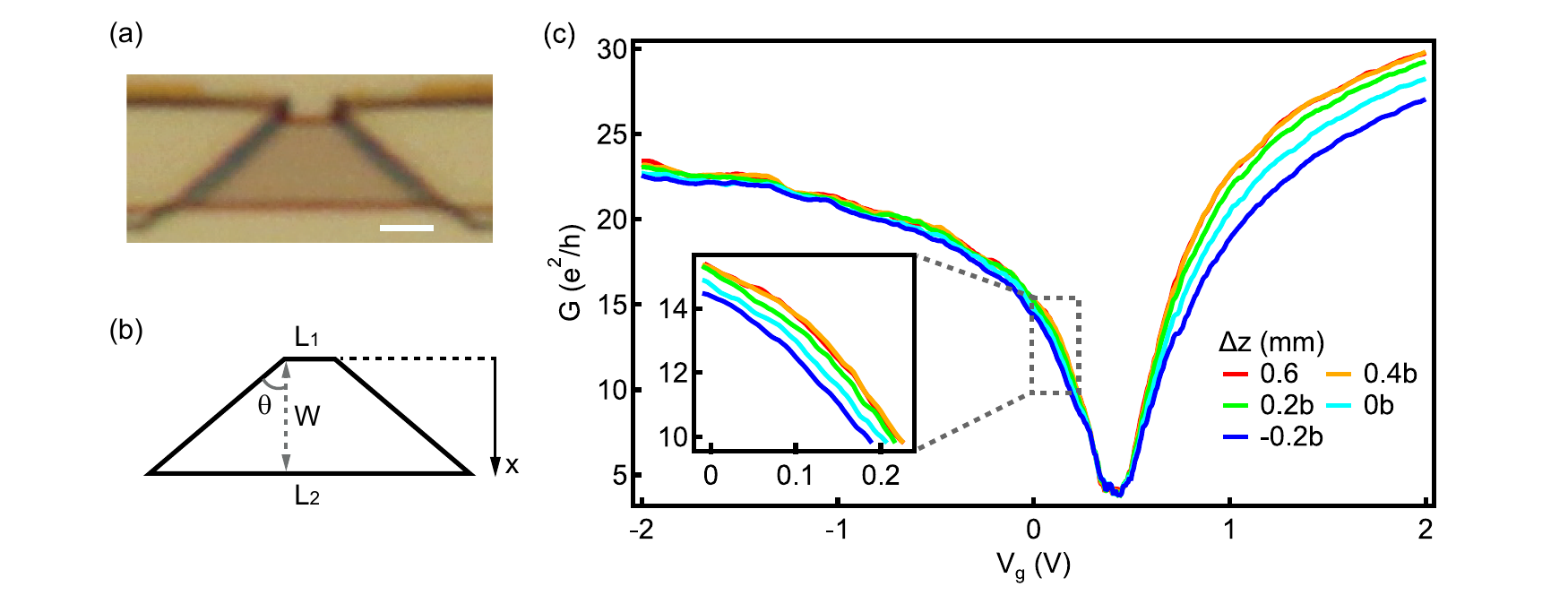}
    \caption{\textbf{(a)} Micrograph of the measured device, scale bar: \SI{1}{\micro\meter}. \textbf{(b)} Sketch of the device for calculating the geometry factor $\alpha$, where $L_1=\SI{1}{\micro\meter}$, $L_2=\SI{6.3}{\micro\meter}$ and $W=\SI{2.25}{\micro\meter}$. \textbf{(c)} Two-terminal differential conductance $G$ plotted as a function of gate voltage $V_\mathrm{g}$ for decreasing $\mathrm{\Delta}z$. ``b'' stands for ``decrease back''. The displacement $\mathrm{\Delta}z$ is defined relative to the mounting position, so negative value does not mean bending in the opposite direction. The inset shows the zoom-in to the hole side.}
    \label{fig:Supp_reversible}
\end{figure}

A micrograph of the device discussed in the main text is shown in Figure~\ref{fig:Supp_reversible}(a). The tapered geometry is intended for other experiments and is not relevant for the conclusions of this study. The dimensions of the device are shown in Figure~\ref{fig:Supp_reversible}(b). Assuming a homogeneous conductivity, the graphene can then be modeled as a set of parallel strips. With this, the geometry factor $\alpha$ for converting the conductance to conductivity can be calculated as:
$$G=\int_{0}^{W}\sigma \frac{dx}{L_{1}+2x\tan \theta },$$ where $\tan\theta =\frac{(L_{2}-L_{1})/2}{W}$, yielding $\sigma =\alpha G$ with $\alpha =\frac{L_{2}-L_{1}}{W\ln(L_{2}/L_{1})}$. Plugging the dimensions of this device, we obtain $\alpha=1.28$.

The two-terminal differential conductance $G$ is also measured for decreasing $\mathrm{\Delta}z$, as shown in Figure~\ref{fig:Supp_reversible}(c). The conductance curves around the CNP become less steep for smaller $\mathrm{\Delta}z$, indicating a decreasing mobility. The change from $\mathrm{\Delta}z=\SI{0.6}{\milli\meter}$ (red curve) to $\mathrm{\Delta}z=\SI{0.4}{\milli\meter}$ (orange curve) is very small, which can be attributed to the mechanical hysteresis of the setup when we turn from increasing $\mathrm{\Delta}z$ to decreasing $\mathrm{\Delta}z$. The observation of decreasing mobility demonstrates that the mobility-increase effect presented in the main text is reversible.

\section{Analysis of the hole side}\label{sec:hole}
\begin{figure}[H]
    \centering
      \includegraphics[width=0.95\columnwidth]{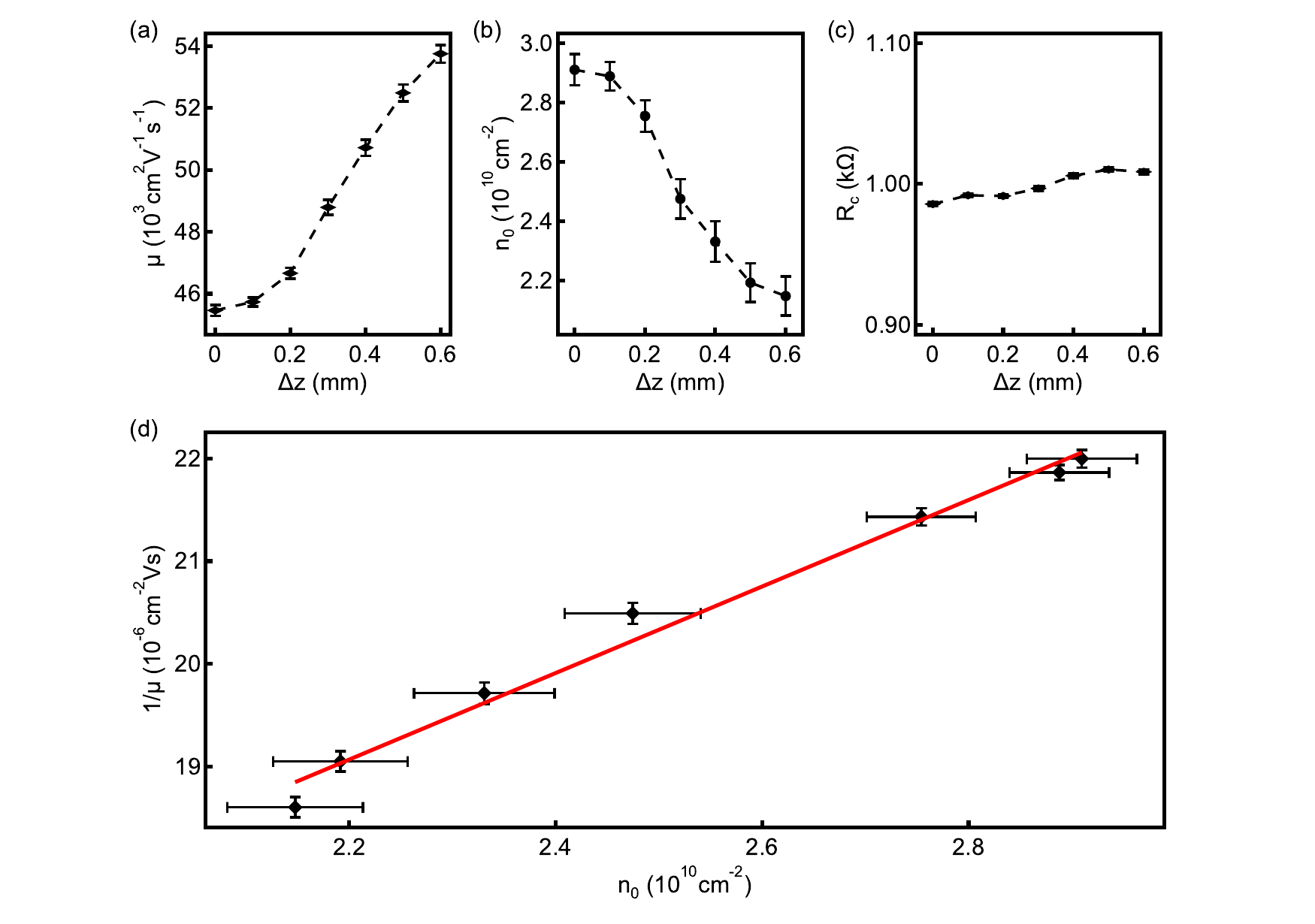}
    \caption{Extracted \textbf{(a)} mobility $\mu$, \textbf{(b)} residual doping $n_{0}$ and \textbf{(c)} contact resistance $R_\mathrm{c}$ (including \SI{\sim350}{\ohm} line resistance) from fitting plotted as a function of $\mathrm{\Delta}z$, respectively. The error bars are the standard errors from fits. The mobility $\mu$ shows a gradual increase while the residual doping $n_{0}$ shows a gradual decrease with increasing $\mathrm{\Delta}z$. The contact resistance $R_\mathrm{c}$ stays essentially constant. \textbf{(d)} Same data as those of \textbf{(a)} and \textbf{(b)} plotted as $1/\mu$ versus $n_{0}$, showing a linear relation with a linear fit to the data, $1/\mu = (0.102\pm0.005) \times (h/e)n_{0}+1/\mu_{0}$, where $\mu_{0}\approx \SI{102000}{\square\centi\meter\per\volt\per\second}$.}
    \label{fig:Hole_side}
\end{figure}

Here we present the results of the fitting on the hole side of Figure~2a in the main text. The extracted $\mu$, $n_{0}$ and $R_\mathrm{c}$ are plotted for different $\mathrm{\Delta}z$ in Figure~\ref{fig:Hole_side}(a-c), respectively. When $\mathrm{\Delta}z$ is increased from 0 to \SI{0.6}{\milli\meter}, the mobility increases gradually from \SI{\sim45000}{\square\centi\meter\per\volt\per\second} to \SI{\sim54000}{\square\centi\meter\per\volt\per\second} while the residual doping decreases gradually from \SI{\sim2.9e10}{\per\square\centi\meter} to \SI{\sim2.1e10}{\per\square\centi\meter\square}. The contact resistance $R_\mathrm{c}$ is \SI{\sim1}{\kilo\ohm} (including \SI{\sim350}{\ohm} line resistance), which is essentially constant and is higher than that on the electron side (\SI{\sim680}{\ohm}) shown in the main text. This is consistent with a p-n junction forming near the contacts which makes the mobility-increase effect less visible on the hole side. The ($\mu$, $n_0$) pairs are plotted as $1/\mu$ versus $n_{0}$ in Figure~\ref{fig:Hole_side}(d). A linear fit yields a slope of $\sim0.102 \times (h/e)$ and $\mu_{0}\approx \SI{102000}{\square\centi\meter\per\volt\per\second}$, both are consistent with the values extracted for the electron side in the main text. 

\section{Weak localization measurements}\label{sec:WL}
\begin{figure}[H]
    \centering
      \includegraphics[width=0.95\columnwidth]{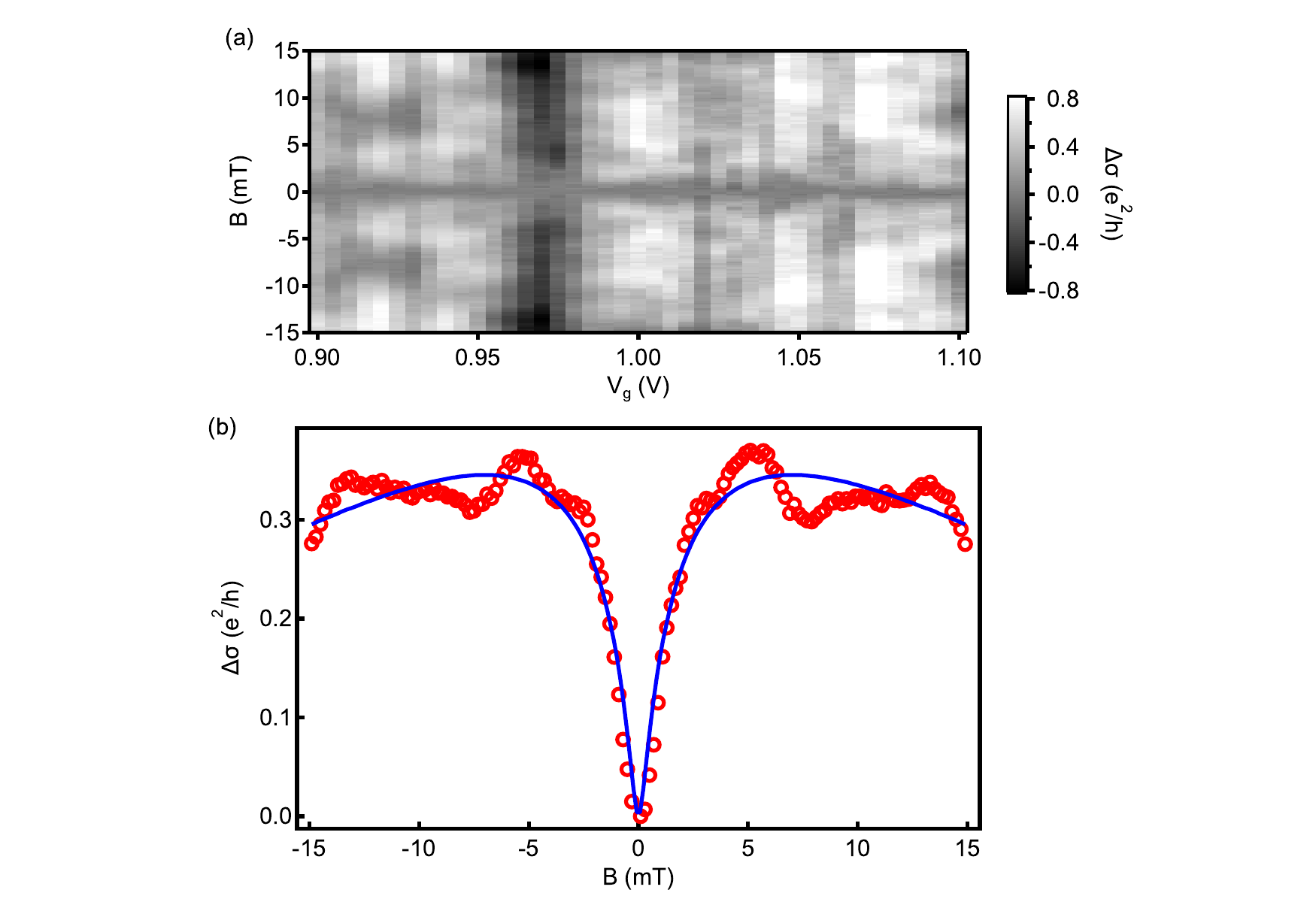}
    \caption{\textbf{(a)} Magnetoconductivity as a function of out-of-plane magnetic field $B$ and gate voltage $V_\mathrm{g}$ measured at $T = \SI{4.2}{\kelvin}$ for $\mathrm{\Delta}z = 0$. A clear feature is observed around $B = 0$. \textbf{(b)} Magnetoconductivity averaged over all traces at different $V_\mathrm{g}$. The red circles represent data that have been ensemble-averaged and the blue line is the fit to the theory of weak localization in graphene.}
    \label{fig:WL}
\end{figure}  

To investigate the characteristic scattering times, we performed weak localization measurements on the device discussed in the main text. Figure~\ref{fig:WL}(a) shows the low-field magnetoconductivity for different $V_\mathrm{g}$ around $V_\mathrm{g} = \SI{1}{\volt}$ for $\mathrm{\Delta}z = 0$. A narrow dip in conductivity is observed around $B = 0$, which is a signature of weak localization. Conductance fluctuations induced by random interference are also visible, which are suppressed by ensemble-averaging over a small $V_\mathrm{g}$ range around the target $V_\mathrm{g}$ value~\cite{Couto2014,Zihlmann2018}. The averaged curve obtained in this way is shown in Figure~\ref{fig:WL}(b) as red circles. We fit the data with the equation from the weak localization theory for graphene~\cite{McCann2006}: $$\mathrm{\Delta} \sigma (B)=\frac{e^{2}}{\pi h}\left [ F\left ( \frac{\tau_\mathrm{B}^{-1}}{\tau_{\phi}^{-1}} \right )- F\left ( \frac{\tau_\mathrm{B}^{-1}}{\tau_{\phi}^{-1}+2\tau_\mathrm{iv}^{-1}} \right )-2F\left ( \frac{\tau_\mathrm{B}^{-1}}{\tau_{\phi}^{-1}+\tau_\mathrm{iv}^{-1}+\tau_\mathrm{*}^{-1}} \right )\right ],$$ where $F(z)=\ln (z) +\Psi(0.5+1/z)$, with $\Psi(x)$ the digamma function, $\tau_\mathrm{B}^{-1}=4eDB/\hbar$ and $D=v_{F}^{2}\tau/2$. The fitting parameters are the phase coherence time $\tau_{\phi}$, the intervalley scattering time $\tau_\mathrm{iv}$ and the intravalley scattering time $\tau_\mathrm{*}$. The elastic scattering time $\tau=\SI{\sim2.6e-13}{\second}$ is determined from the carrier mobility and is not a fitting parameter here. The fitted curve is plotted as blue solid line in Figure~\ref{fig:WL}(b) with the results $\tau_{\phi}=\SI[separate-uncertainty = true]{9.9(5)e-12}{\second}$, $\tau_\mathrm{iv}=\SI[separate-uncertainty = true]{5.8(6)e-12}{\second}$ and $\tau_\mathrm{*}=\SI[separate-uncertainty = true]{2.8(2)e-13}{\second}$. As already stated in the main text, $\tau_\mathrm{iv}\gg\tau$ while $\tau_\mathrm{*}$ is comparable to $\tau$, implying that intravalley scattering due to local stain fluctuations rather than charged impurities is the process limiting the mobility~\cite{Couto2014}.


\section{Raman Measurements}\label{sec:Raman}
After transport measurements at low temperature, we perform Raman measurements at room temperature to determine the strain if the device is still intact. The Raman setup is separate from the transport setup and sits in ambient conditions~\cite{Wang2019a}. We use the commercially available confocal Raman system WiTec alpha300. All Raman spectra were acquired using a linearly polarized laser at a wavelength of $\SI{532}{\nano\meter}$ and a power of $\SI{1}{\milli\watt}$. The laser spot size is $\SI{\sim500}{\nano\meter}$ and the grating of the spectrometer is 600 grooves/mm.

\section{Second device with mobility-increase effect}\label{sec:second}
\begin{figure}[H]
    \centering
      \includegraphics[width=\columnwidth]{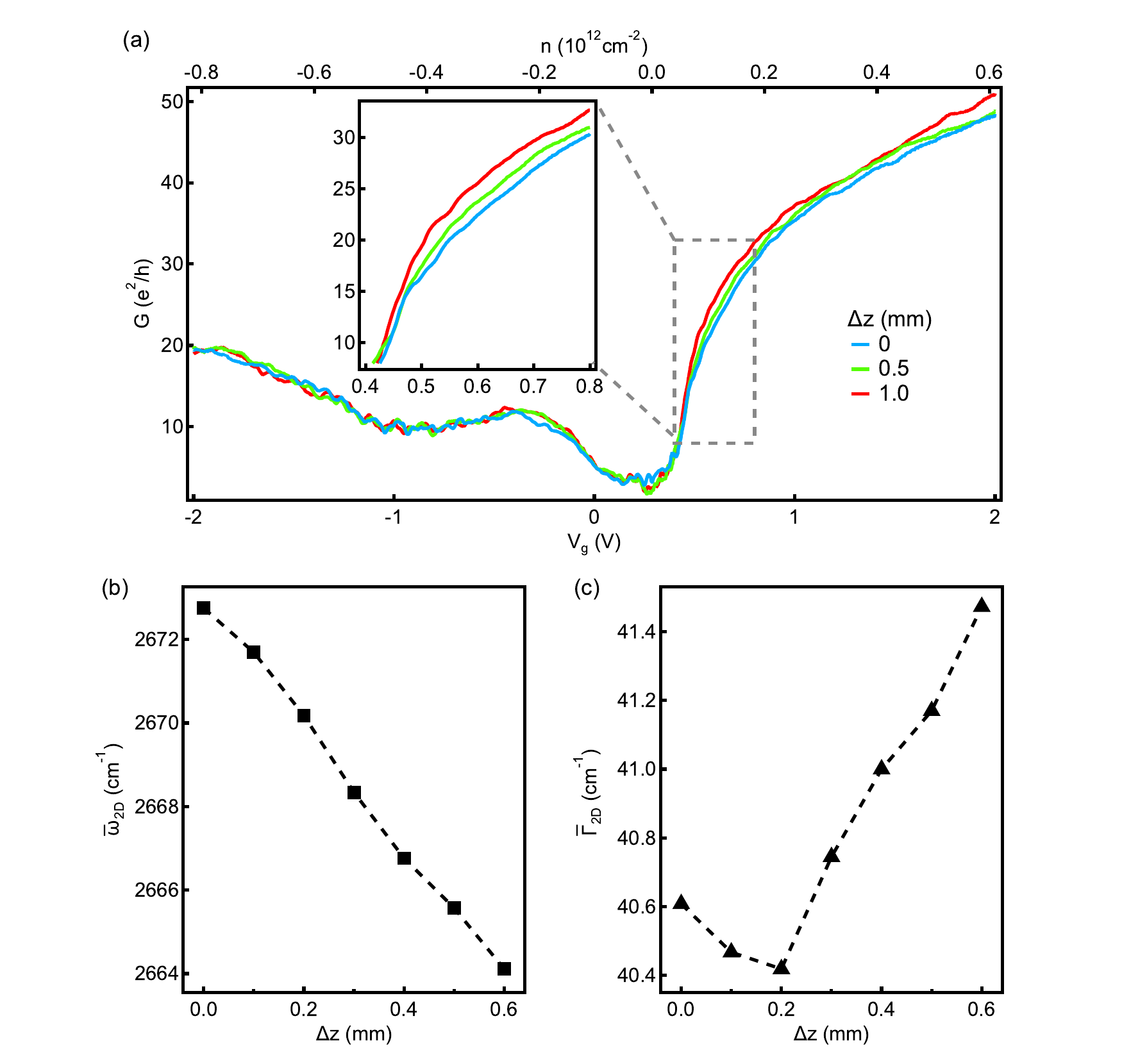}
    \caption{\textbf{(a)}Two-terminal differential conductance $G$ plotted as a function of gate voltage $V_\mathrm{g}$ for different $\mathrm{\Delta}z$. The corresponding carrier density is shown on the top axis. \textbf{(b)} Averaged center frequency $\bar{\omega}_{2D}$ and \textbf{(c)} linewidth $\bar{\Gamma}_{2D}$ of the Raman 2D peak plotted as a function of $\mathrm{\Delta}z$ for the same device.}
    \label{fig:device2}
\end{figure} 

The electron mobility of this device is around \SI{70000}{\square\centi\meter\per\volt\per\second}, which shows a slight increase with increasing $\mathrm{\Delta}z$, as plotted in Figure~\ref{fig:device2}(a). The visibility of this effect on the hole side is suppressed by the higher contact resistance on this side and the additional conductance minimum at $V_\mathrm{g}\approx\SI{1}{\volt}$, which originates from a double moir\'e superlattice effect in encapsulated graphene when both the top and the bottom hBN lattices are aligned to the graphene lattice~\cite{Wang2019}. In Raman measurements, the mean value of the center frequency $\bar{\omega}_\mathrm{2D}$ and the linewidth $\bar{\Gamma}_\mathrm{2D}$ of the 2D peak averaged over the whole device are plotted as a function of $\mathrm{\Delta}z$ in Figure~\ref{fig:device2}(b,c), respectively. The value of $\bar{\omega}_\mathrm{2D}$ shows a linear decrease with increasing $\mathrm{\Delta}z$, indicating an increasing average strain. For $\bar{\Gamma}_\mathrm{2D}$, it decreases first with $\mathrm{\Delta}z$, implying reducing of the local strain fluctuations due to increasing average strain. The increase of $\bar{\Gamma}_\mathrm{2D}$ appears when the externally applied strain induced broadening of the 2D peak dominates. The overall large values of $\bar{\Gamma}_\mathrm{2D}$ come from the superlattice effect~\cite{Eckmann2013}, which is also observed in transport measurements as shown in Figure~\ref{fig:device2}(a).

\section{Example device without strain induced mobility increase}
\label{sec:no_change}
\begin{figure}[H]
    \centering
      \includegraphics[width=\columnwidth]{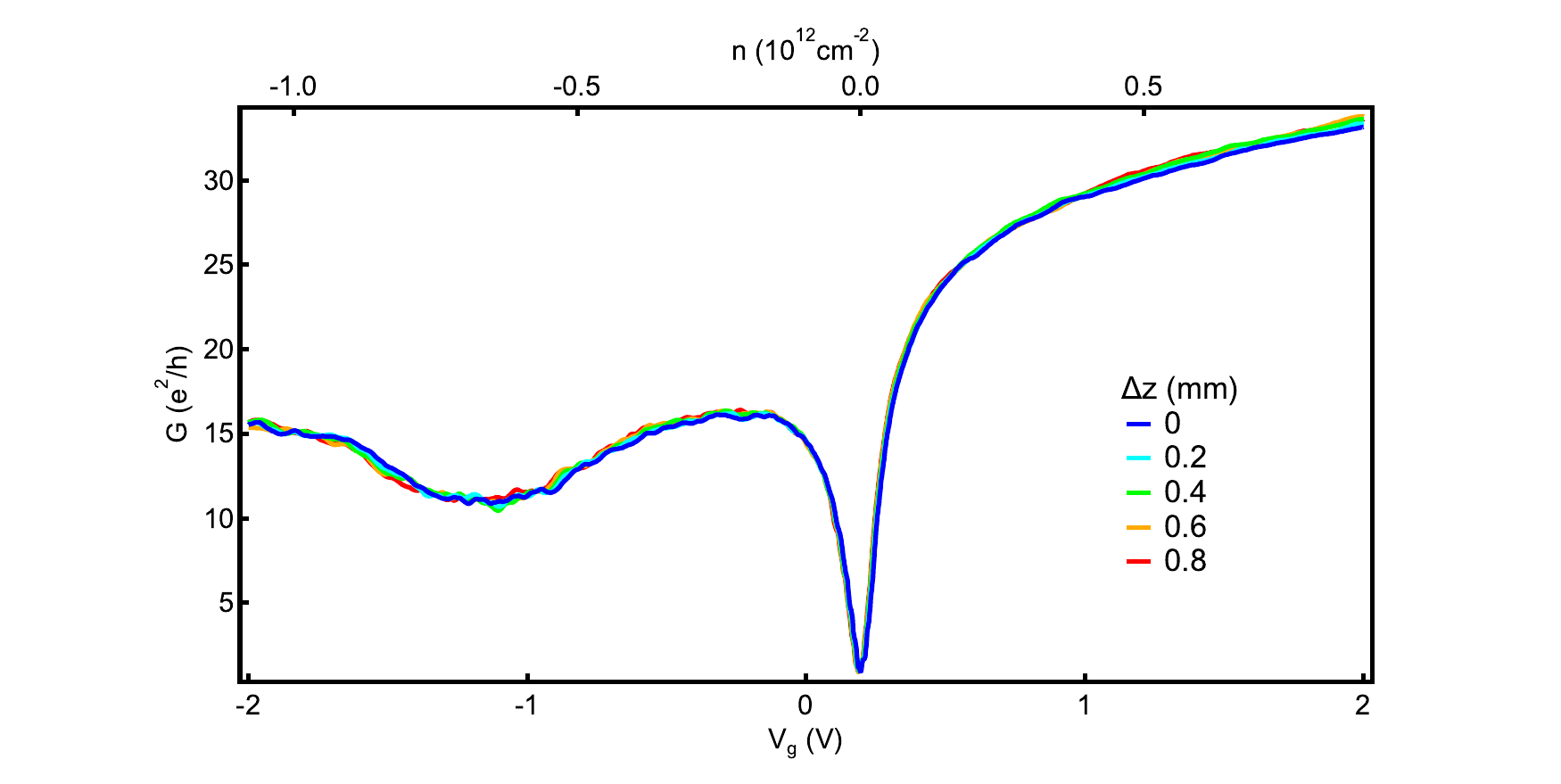}
    \caption{Two-terminal differential conductance $G$ measured as a function of gate voltage $V_\mathrm{g}$ for different $\mathrm{\Delta}z$. The corresponding carrier density is shown on the top axis.}
    \label{fig:CuBJ13_J5}
\end{figure} 

Here we present an example of the devices which do not show the mobility-increase effect. The mobility values of such devices are all higher than \SI{80000}{\square\centi\meter\per\volt\per\second}. As shown in Figure~\ref{fig:CuBJ13_J5}, the mobility of this device does not increase with increasing $\mathrm{\Delta}z$. The mobility value is above \SI{100000}{\square\centi\meter\per\volt\per\second}, which is beyond the highest value of the studied range in Ref.~\cite{Couto2014}. The absence of the mobility-increase effect in this device suggests that either the strain fluctuations cannot be fully reduced by increasing the average strain or another scattering mechanism becomes relevant for ultra high mobility devices. 

\bibliography{MobilityChange}